\documentclass[nofootinbib,showpacs,preprintnumbers,amsmath,amssymb]{revtex4}
\usepackage{epsfig}

\begin{document}

\title{Impact of double-logarithmic electroweak radiative corrections
on the non-singlet structure functions at small $x$ }

\vspace*{0.3 cm}

\author{B.I.~Ermolaev}
\affiliation{Ioffe Physico-Technical Institute, 194021
 St.Petersburg, Russia}

\author{S.I.~Troyan}
\affiliation{St.Petersburg Institute of Nuclear Physics, 188300
Gatchina, Russia}

\begin{abstract}
In the QCD context, the non-singlet structure functions of $u$ and
$d$ -quarks are identical, save the initial quark densities.
Electro-weak radiative corrections, being flavor-dependent, bring
further difference between the non-singlets. This difference is
calculated in the double-logarithmic approximation and estimated
numerically.
\end{abstract}

\maketitle

\section{Introduction}
Double-logarithmic (DL) contributions  were discovered in
Ref.~\cite{sud} in the QED context and since  that have become  a
popular object of theoretical investigations. On one hand, DL
terms are among the most sizable radiative corrections in each
order of the field theories at high energies. On the other hand,
the ways to select the Feynman graphs yielding DL terms, the means
to calculate DL contributions and the methods of all-order
summations first developed in Ref.~\cite{ggfl} converted earlier
examples of DL calculations into the regular technique that allows
to account for DL radiative  corrections  in a quite efficient and
simple way. With certain technical modifications, especially
non-trivial for inelastic processes, the general prescriptions of
calculating DL asymptotics elaborated in Ref.~\cite{ggfl} were
generalized to QCD and the Standard Model of the electro-weak
interactions at TeV energies where the total energy $\sqrt{s} \gg
M_{W,Z}$. As for the electro-weak (EW) double-logarithms, quite
often in the literature they are accounted in fixed orders in the
EW couplings. Ref.~\cite{flmm} proved the exponentiation of the
soft EW DL contributions.
% and thereby initiated activity on systematic
%calculations of the DL asymptotics of electro-weak reactions.
Such an exponentiation takes place for electro-weak reactions in
the hard kinematics. The more involved Regge kinematics was
studied in Refs.~\cite{bar,egtew}. One of the most essential
difference between EW and other DL calculations is the fact that
the gauge symmetry of the EW interactions is partly broken and the
set of the EW bosons includes the massless (photons) and massive
(W,Z) particles. The DL contributions involving soft photons are
infrared-divergent and are regulated with the infrared cut-off
$\mu$ exactly as in QED. The value of $\mu$ is fixed in final
formulas with physical considerations. DL contributions involving
soft $W,Z$-bosons are infrared-stable and contain, instead of
$\mu$, masses $M_w,~M_Z$ of the involved bosons. The difference
between $M_W$ and $M_Z$ can be neglected with the DL accuracy. It
makes possible to use the second cut-off, $M$ (with $M \geqslant
M_W \approx M_Z$) instead of $M_w,~M_Z$ in the DL contributions
from virtual $W$ and $Z$ -bosons. This approximation considerably
simplifies all-order summations of EW double-logs. Another
interesting topic is the interplay between the QCD and EW
double-logarithmic contributions. In particular, it was stated in
Ref.~\cite{ross} that the impact of the first-loop EW
double-logarithmic terms on hadronic reactions ($\equiv$ EW
impact) can be as large as $10 \%$ at energies $\sqrt{s} \sim
500~$GeV;
%(in the first version of Ref.~\cite{ross} this estimate was $20 \%$)
the role of sub-leading contributions was discussed in
Ref.~\cite{ross1}.

In the present paper we examine the EW impact first for the $2 \to
2$ quark scattering amplitudes and then for the inclusive cross
sections. We show that the EW impact on the elastic QCD scattering
amplitudes calculated in the first loop is smaller than the
estimates made in Ref.~\cite{ross}: the impact is
%$\approx 2.7 \%$ at $\sqrt{s} \sim 500~$GeV, then it reaches
$\approx 3.5 \%$ at $\sqrt{s}= 1~$TeV.
%It immediately leads to the EW impact $\approx \leqq 7~\%$ for the exclusive cross
%sections at $\sqrt{s} \leqq 10~$TeV.
The total resummation of DL contributions to the elastic
scattering amplitudes increases the EW impact compared to the
first-loop impact: the impact is $10~\%$ at $\sqrt{s} = 1~$TeV and
growing fast with $\sqrt{s}$ reaches $30 \%$ at  $\sqrt{s} =
10~$TeV. The EW impact on the amplitudes of the inelastic $2 \to 2
+ n$ -scattering of quarks can be estimated similarly. The
explicit expressions for such amplitudes in QCD were obtained in
Ref.~\cite{el} and the generalization to the electroweak processes
can be found in Ref.~\cite{eot}. On the other hand, the EW
radiative corrections depend on the flavors of the involved
quarks, so accounting for these EW corrections in DLA together
with the QCD background can bring qualitatively new phenomena. For
example, let us consider the flavor non-singlet contributions to
the structure functions $F_1$ and $g_1$ of Deep-Inelastic
Scattering (DIS), i.e. the flavor-depended contributions to the
inclusive cross sections of the DIS. They are often addressed as
the non-singlet structure functions $ f^{(\pm)}(x, Q^2)$. As is
well-known, the expressions for $f^{(\pm)}(x, Q^2)$ include the
coefficient functions (to describe the $x$ -evolution), anomalous
dimensions (to describe the $Q^2$- evolution) and the initial
quark densities $\delta q$, with $\delta q = \delta u,~\delta d$.
When calculated in the QCD framework, $f^{(\pm)}(x, Q^2)$ for the
$u$- quark and $d$- quark coincide, save difference between $e^2_u
\delta u$ and $e^2_d \delta d$: the quark-gluon interactions do
not depend of flavors of the quarks. Electroweak corrections to
$f^{(\pm)}$ bring more difference: they cause difference in the
$x$ and $Q^2$ -evolutions of the initial quarks and split
$f^{(\pm)}$ into $f^{(\pm)}_u$ and $f^{(\pm)}_d$ (the subscripts
$u,~d$ label the initial quark flavors). The difference in the
evolutions of $u$ and $d$ -quarks means that $f^{(\pm)}_u \neq
f^{(\pm)}_d$ even if $e^2_u \delta u = e^2_d \delta d$. Impact of
the electromagnetic $\sim O(\alpha)$ corrections was studied in
Ref.~\cite{mrst} where DGLAP evolution equation\cite{dglap} was
used for accounting for the QCD corrections. However, DGLAP does
not include resummation of the DL terms $\sim
\alpha_s^k\ln^{2k}(1/x)$ and the single-logarithmic (SL) terms
$\sim \alpha_s^k\ln^k(1/x)$. The point is that DGLAP was
originally suggested for operating within the region of large $x$
where the DL and SL contributions could easily be neglected in
higher loops. Accounting for them
 to all orders in $\alpha_s$ becomes
necessary in the small-$x$ region. DGLAP lacks the resummation, so
the extrapolation of DGLAP into the small-$x$ region involves
introducing the singular fits for $\delta q$ with many
phenomenological parameters (see e.g. Ref.~\cite{a}) but suggests
no theoretical explanations why $\delta u$ and $\delta d$ should
be singular.
%Exploiting such fits causes the steep rise of $f_{NS}$ at small
%$x$ and brings a satisfactory agreement with experimental data
% but the price is .
In fact, the only role of the singular terms in the fits is  to
mimic the total resummation of the leading logarithms of $x$(see
Ref.~\cite{egtinp} for more detail). When the resummation is taken
into account, the singular factors should be dropped and therefore
the fits can be simplified. On the other hand, the total
resummation of the EW DL contributions to $f^{(\pm)}$
%together with the QCD DL and SL contributions
makes possible to estimate their impact on the small-$x$ behavior
of the non-singlets. In doing so, we follow the approach of
Refs.~\cite{egtns,bar,egtew}. Through the paper we neglect the
running effects for the EW couplings.

The present paper is organized as follows: In Sect.~II we consider
 the EW impact on the simplest and at the same time basic exclusive QCD process:
 the quark-antiquark annihilation in the hard kinematics. After that we
 study the EW impact on the inclusive cross sections in QCD.  One can easily
 anticipate that the
 EW impact on the singlet components of the DIS structure functions should
 be very small: the main contributions to the singlets comes from the gluon ladder
 graphs and gluons do not participate in the EW interactions. So, we consider
 the EW impact on the non-singlet structure functions where the main contributions
 come from the quark ladders. In
Sect.~III we remind the results of Ref.~\cite{egtns} for the
non-singlet structure function $f^{(\pm)}$ in QCD. This expression
is the solution of the Infrared Evolution Equation (IREE). From
pedagogical reasons, in Sect.~IV we first extend the QCD results
for $f^{(\pm)}$, including the electromagnetic DL corrections. The
system of IREE were all EW DL corrections are taken into account
is obtained in Sect.~V.  In contrast to QCD, the evolution
equations for $f_{u,d}$ involve four anomalous dimensions. The
IREE for them are composed in Sect.~VI. Besides, in order to solve
the IREE for $f_{u,d}$, auxiliary amplitudes are obtained in
Sect.~VII. It makes possible to obtain  explicit expressions for
$f_{u,d}$ first in the Mellin space in Sect.~VIII and then in the
conventional form in Sect.~IX.   In Sect.~X we consider the
small-$x$ asymptotics of the non-singlet structure functions and
estimate the impact of the EW corrections on the non-singlet
intercepts. Finally, Sect.~XI is for concluding remarks.

\section{DL electroweak corrections to amplitudes of $2\to2$ scattering in the hard kinematics}

In order to estimate the impact of EW DL contributions on
exclusive processes in QCD, we consider the $2 \to 2$ -scattering
in the hard kinematics, i.e. in the kinematics where all
Mandelstamm variables are of the same order:
\begin{equation}\label{hardkin}
s \approx -t \approx -u.
\end{equation}
The impact of EW double logarithms on amplitudes of gluon
scattering is surely less than the one for quarks, so we consider
the scattering amplitude $A$ of the annihilation of the
quark-antiquark pair $q~\bar{q}$ into another quark-antiquark pair
$q'~\bar{q}'$, assuming that the flavors of $q$ and $q'$ are
different. As is known, double-logarithms are leading
contributions among radiative corrections to this process.
Calculating the DL radiative corrections to the scattering
amplitudes of $2 \to 2$ -processes in the hard kinematics is
rather simple from the technical point of view because the most
difficult, ladder Feynman graphs do not yield DL contributions in
the kinematics (\ref{hardkin}). For calculations with the DL
accuracy
 in the hard kinematics it is convenient to use the
Coulomb gauge where DL contributions arrive from the self-energy
graphs only, which simplifies the calculations a lot. It is easy
to check that this remarkable feature does not take place in the
Regge kinematics that we consider in the next Sects. When both QCD
and EW double-logarithms are accounted for, the first-loop
contribution $S$ is the sum of $S_r$, with $r =
q,q',\bar{q},\bar{q}'$:
\begin{equation}\label{psiq}
S = \sum_r S_r = 2S_u + 2S_d
\end{equation}
where the subscripts $u,d$ refer to the up- and down- quarks
respectively. In the QCD context and when $\alpha_s$ is fixed,
\begin{equation}\label{psirqcd}
S_u  = S_d  = S_{QCD} = \frac{1}{8\pi}\alpha_s C_F \ln^2(s/\mu^2),
\end{equation}
with $\mu$ being the infrared cut-off. When both QCD and EW double
logarithms are taken into account,
\begin{equation}\label{psir}
S_r = \frac{1}{8\pi}\Big[(\alpha_s C_F+ \alpha Q^2_r)
\ln^2(s/\mu^2) + \frac{\alpha}{\sin^2 \theta_W}\Big(C'_F +
 \frac{Y^2}{4}\tan^2 \theta_W \Big)\ln^2(s/M^2) -\alpha Q^2_r
\ln^2(s/M^2)\Big]
\end{equation}
where $\mu$ is the common infrared cut-off for gluons and photons.
In Eqs.~(\ref{psir},\ref{psirqcd}) $\theta_W$ is the Weinberg
angle and $C_F,C'_F$ defined as $(N^2 -1)/(2N)$ for the groups
$SU(3)$ and $SU(2)$ respectively so that $C_F = 4/3,~C'_F = 3/4$.
We also use the conventional notations $T, Y$ and $Q_r$ for the
isospin, hypercharge and electric charge of quarks. They obey the
standard relation $Q_r = T_{3~r} + Y/2$. We have also used and
will keep through the paper the approximation $M_Z \approx M_W =
M$.
 When $\alpha_s$ is running, its argument in the hard
kinematics is $k^2_{\perp}$, so the QCD contribution
$(1/2)\alpha_s C_F \ln^2 (s/\mu^2)$ in
Eqs.~(\ref{psirqcd},\ref{psir}) should be replaced by
\begin{equation}\label{alfasrun}
\int_{\mu^2}^{s} \frac{d
k^2_{\perp}}{k^2_{\perp}}\alpha_s(k^2_{\perp}) C_F  \ln
(s/k^2_{\perp})= \frac{C_F}{b} \int_{\mu^2}^{s} \frac{d
k^2_{\perp}}{k^2_{\perp}} \frac{\ln
(s/k^2_{\perp})}{\ln(k^2_{\perp}/\Lambda^2_{QCD})} = \frac{C_F}{b}
\Big[ \ln \Big(\frac{\ln
(s/\Lambda^2_{QCD})}{\ln(\mu^2/\Lambda^2_{QCD})}\Big)
\ln(s/\Lambda^2_{QCD}) - \ln(s/\mu^2)\Big]
\end{equation}
where $b$ is the Gell-Mann-Low function in LO. Accounting for the
total resummation of DL contributions leads to exponentiation of
the first-loop contribution $S$. It converts the Born amplitude
$A_{Born}$ into
\begin{equation}\label{amplhard}
A =  A_{Born} e^{-S}.
\end{equation}
Let us first estimate the impact of EW double-logarithms in the
first-loop order. Defining this impact as
\begin{equation}\label{flimp}
\epsilon^{(1)} = \frac{S_r - S_{QCD}}{S_{QCD}},
\end{equation}
then putting $Q_r = 2/3$ and estimating $
%\alpha_s = 0.12,~
\mu =1~$GeV, $M = 90~$GeV, we obtain that $\epsilon^{(1)}$ slowly
grows with the total energy $\sqrt{s}$:
%\begin{equation}\label{imp1}
$\epsilon^{(1)}  \lesssim 3 \%$
%\end{equation}
at $\sqrt{s} \lesssim 10^3~$GeV; then $\epsilon^{(1)}$ reaches $10
\%$ at $\sqrt{s} \sim 10^5~$GeV. Basically, $\mu$ is not fixed; in
formulas for total cross sections it is usually replaced by
suitable physical quantities like energy resolutions, minimal
transverse momenta, etc. Here we use the estimate $\mu =1~$GeV
because  it is the scale typical for QCD and then it coincides
with the choice of $\mu$ for the non-singlet structure functions
which we will consider in the next Sects. (see Ref.~\cite{egtns}
for detail). Therefore, the first-loop estimate for the EW impact
to the amplitudes of the $2 \to 2$ quark scattering is small for
$\sqrt{s} \lesssim 10^3~$GeV. It immediately leads to the estimate
$\lesssim 7~\%$ for the EW impacts on the exclusive cross
sections.  However, the situation looks more optimistic for the
case of the total resummation of the double-logarithms: Let us
define, similarly to Eq.~(\ref{flimp}), the EW impact
$\epsilon_{hard}$ for the amplitude $A$ in Eq.~(\ref{amplhard}):
\begin{equation}\label{imphard}
\epsilon_{hard} = \frac{|A - A_{QCD}|}{A_{QCD}}
\end{equation}
where the QCD amplitude $A_{QCD}$ is given by Eq.~(\ref{amplhard})
with $S = 4S_{QCD}$. It is easy to obtain that $\epsilon_{hard}$
grows with $s$ much faster than $\epsilon^{(1)}$, achieving
$\approx 10 \%$ at $\sqrt{s} \approx 10^3~$GeV and exceeds $30 \%$
at $\sqrt{s} \approx 10^4~$GeV.

\section{Non-singlet structure function at small $x$ in the QCD framework}

The term "non-singlet structure functions" stands for
flavor-dependent contributions to DIS structure functions.
Usually, DIS structure functions are calculated with using the
DGLAP evolution equations. As is known, DGLAP accounts for
logarithms of $Q^2$ to all orders in the QCD coupling $\alpha_s$
and at the same time lacks the total resummation of Double- and
Single logarithms (DL and SL respectively) of $x$. Such
contributions are important at small $x$. The total summation of
them, including the running coupling effects, was
 performed in Refs.~\cite{egtns} with composing and solving the Infra-Red Evolution Equations
 (IREE). We will use this approach in the present paper in order to account
 for EW DL contributions, so we briefly remind below of the QCD
 results for the non-singlet structure functions. In order to make clear the fact that
 we discuss in this section
 only the QCD content of the non-singlet structure function, we will
 use the subscript "QCD" where it is necessary. Usually, notations (like $f_{NS}$) for
 the non-singlet structure functions bear the subscript $"NS"$ (and the subscript $"S"$
 is reserved for the singlet structure functions) but as through
 the paper we discuss the non-singlets only, we do not write the subscript
 $"NS"$. We denote $f^{(+)}$ the non-singlet contribution  to the
 unpolarized structure function $F_1$ and use the notation
 $f^{(-)}$  for the non-singlet contribution  to the
 spin structure function $g_1$. As is known,
 the latter coincides with the structure function $f_3$.
 Technically, it is convenient
 to introduce
 the forward Compton
amplitudes $T^{(\pm)}(s,Q^2)$ related with $f^{(\pm)}$ by the
Optical theorem:
\begin{equation}\label{compt}
    f^{(\pm)}(x, Q^2) = \frac{1}{\pi} \Im T^{(\pm)}(s, Q^2)
\end{equation}
where we have used the standard notations: $q$ is the momentum of
the incoming virtual photon, $p$  is the incoming quark momentum,
$Q^2 = - q^2$, $x = Q^2/2pq$, $s = (p + q)^2 \approx 2pq$ when $x
 \ll 1$. The superscripts $"\pm"$ in Eq.~(\ref{compt}) manifest that
amplitudes $T^{(\pm)}$ have the signatures $\pm$. It means  that
they are defined as  follows:
\begin{equation}\label{sign}
T^{(\pm)}  = \frac12 [ T(s,Q^2)\pm T(-s,Q^2) ]~.
\end{equation}
Using  the signature amplitudes at high energies is absolutely
necessary from the point of view of the  phenomenological Regge
theory and  at the same time it is convenient technically (see
e.g. Ref.~\cite{egtns} for detail).
 Accounting for the summation of the  DL contributions
 $\sim (\alpha_s \ln^2(1/x))^k,~(k=1,...)$ makes necessary introducing an
 infrared cut-off  $\mu$. For the sake of simplicity we identify it with the
 starting point of  the $Q^2$-evolution, though it is not
 necessary. Therefore, both $T^{(\pm)}$ and $f^{(\pm)}$ depend on
$\mu$ as well.
 %As it is, $f^{(\pm)}$ evolve the initial quark densities $\delta q$
%from the region $x \lesssim 1,~ Q^2 = \mu^2 \approx 1$GeV$^2$ to
%the region $x \ll 1,~ Q^2 \gg \mu^2$, so both $f_{QCD}^{NS}$ and
%$T_{QCD}(s, Q^2)$ depend also on the $\mu$. In the present paper
%we focus on studying $f^{NS}$ at $x \ll 1$. The evolution
%equations for $f^{NS}$ contain convolutions.
%In order to simplify  them,
It is convenient (see Ref.~\cite{egtns} for detail) to use an
integral transform to represent $f^{(\pm)}$ and $T^{(\pm)}$.  The
Regge pole theory suggests that it should be the Sommerfeld-Watson
transform. At $s\to\infty$ one can use its asymptotic form that
looks quite similarly to the Mellin transform:
\begin{equation}
\label{mellin} T^{(\pm)} = \int_{-\imath \infty }^{\imath
\infty}\frac{d\omega}{2\pi \imath}
\Big(\frac{s}{\mu^2}\Big)^{\omega} \xi^{(\pm)}(\omega)
F^{(\pm)}(\omega, y)
\end{equation}
where the signature factors
\begin{equation}\label{signfact}
\xi^{(\pm)} = [ e^{-\imath \pi \omega}\pm 1 ] / 2 \approx [ 1 \pm
 1 - \imath \pi \omega ] /2~.
\end{equation}
 As Eq.~(\ref{mellin}) partly coincides with the
standard Mellin transform, it is often addressed as the Mellin
transform and we will do the same through this paper.
Nevertheless, we will use the transform inverse to
Eq.~(\ref{mellin}) in its proper form:
\begin{equation}\label{invmellin}
F^{(\pm)}(\omega, y) =  \frac{2}{\pi \omega} \int_{0}^{\infty} d
\rho e^{-\omega \rho} \Im  T^{(\pm)}(\rho,y)
\end{equation}
where we have introduced two new convenient variables $\rho  =
\ln(s/\mu^2)$ and $y = \ln(Q^2/\mu^2)$. Obviously,
Eq.~(\ref{invmellin}) does not coincide with  the standard Mellin
transform.

Eqs.~(\ref{compt},\ref{mellin}) read that
\begin{equation}
\label{ft} f^{(\pm)}= (1/2)\int_{-\imath \infty }^{\imath
\infty}\frac{d\omega}{2\pi \imath}
\Big(\frac{s}{\mu^2}\Big)^{\omega} \omega F^{(\pm)}(\omega, y)~.
\end{equation}
Evolving amplitudes $T^{(\pm)}$ with respect to $\mu$ allows  one
to compose IREE for them. It was shown in Ref.~\cite{egtns} that
in the QCD framework the forward Compton amplitudes $T^{(\pm)}$
obey the following equation:
\begin{equation}\label{iree}
T^{(\pm)}= T^{(\pm)}_{Born}  + M_0^{(\pm)} \otimes T^{(\pm)}
\end{equation}
where $T^{(\pm)}_{Born}$ is $T^{(\pm)}$ in the Born approximation,
$M_0^{(\pm)}$ are amplitudes of the forward quark-quark
scattering. They should be calculated independently. After
differentiating Eq.~(\ref{iree}) with respect to $\mu$ and
applying the Mellin transform, Eq.~(\ref{iree})  converts into the
following equation in terms of $F^{(\pm)}_{QCD}(\omega, y)$:
\begin{equation}\label{eqfqcd}
(\omega + \partial/ \partial y)F^{(\pm)}_{QCD}  = [1 + \omega/2]
H^{(\pm)}_{QCD}(\omega) F^{(\pm)}_{QCD} ~.
\end{equation}
The Born term $T^{(\pm)}_{Born}$ does not depend on $\mu$ and
vanishes after the differentiation. The term $\omega/2$ in
Eq.~(\ref{eqfqcd}) describes the single-logarithmic contribution.
As our aim is studying DL contributions, we will neglect such SL
contributions through the paper, though we will keep $\alpha_s$
running.
 $H^{(\pm)}_{QCD}(\omega)$ in Eq.~(\ref{eqfqcd}) are related to
 amplitudes $M_0^{(\pm)}$ through the Mellin transform. They are new
anomalous dimensions. They include the total resummation of DL and
SL QCD contributions. IREE for $H^{(\pm)}_{QCD}$ obtained in
Ref.~\cite{egtns}. When the SL terms that do not contribute to
$\alpha_s$ are neglected, the IREE for $H^{(\pm)}_{QCD}$ is
\begin{equation}\label{eqhqcd}
 \omega H^{(\pm)}_{QCD} = \frac{b^{(\pm)}_{QCD}}{8\pi^2} +  \big(H^{(\pm)}_{QCD}\big)^2
\end{equation}
where
\begin{equation}\label{badqcd}
b^{(\pm)}_{QCD} = a_{QCD} + D^{(\pm)}_{QCD}~,
\end{equation}
with
\begin{equation}\label{a} a_{QCD} = 4\pi A(\omega)C_F~,\qquad A(\omega) =
\frac{1}{b} \Big[\frac{\eta}{\eta^2 + \pi^2} - \int_0^{\infty}
\frac{d \rho e^{-\omega \rho}}{(\rho + \eta)^2 + \pi^2} \Big]~
\end{equation}
and
\begin{equation}\label{dint}
D^{(\pm)}_{QCD}(\omega)  = \big(-\frac{C_F}{2N}\Big)(-4)
\int_0^{\infty} d \rho  e^{-\omega \rho}~ \Re[\alpha_s(s)\mp
\alpha_s(-s)] \int_{\mu^2}^{s}  \frac{d
k^2_{\perp}}{k^2_{\perp}}\alpha_s(k^2_{\perp})~.
\end{equation}
Performing integration over $k^2_{\perp}$ in Eq.~(\ref{dint}), we
obtain the following expression for $D^{(\pm)}(\omega)_{QCD}$:
\begin{equation}\label{d}
D^{(\pm)}_{QCD}(\omega)  = \frac{2C_F}{b^2 N} \int_0^{\infty} d
\rho e^{-\omega \rho} \ln \big( \frac{\rho + \eta}{\eta}\big)
\Big[ \frac{\rho + \eta}{(\rho + \eta)^2 + \pi^2} \mp
\frac{1}{\rho + \eta}\Big] ~.
\end{equation}
   In Eqs.~(\ref{a},\ref{d}) $\rho  = \ln(s/\mu^2),~\eta =
\ln(\mu^2/\Lambda^2_{QCD})$~, and we have used the standard
notations: $C_F = (N^2 - 1)/2N = 4/3$ and $b$ is the first
coefficient of the Gell-Mann-Low function.

Eqs.~(\ref{iree}-\ref{d}) were obtained and discussed in detail in
Ref.~\cite{egtns}, so in the present paper we do not derive them.
Instead, we show in next Sects how to extend the QCD results,
Eqs.~(\ref{iree}-\ref{d}), to the Standard Model of electroweak
interactions. Nevertheless, let us briefly comment them. The term
$a_{QCD}/(\omega )$ in Eqs.~(\ref{eqfqcd},\ref{badqcd}) is the
Born contribution to the amplitudes of the forward quark-quark
scattering, so that $A(\omega)$ is related to $\alpha_s$ through
the Mellin transform of Eq.~(\ref{invmellin}). In contrast, the
Born contribution is absent in Eq.~(\ref{eqfqcd}) because it does
not depend on $\mu$ and therefore vanishes under differentiation
over $\mu$. The second term, $D(\omega)$ in Eq.~(\ref{badqcd})
represents the approximative DL contribution of non-ladder Feynman
graphs\footnote{Through this paper we use the Feynman gauge.} when
the $s$ and $u$ -channel gluons with small transverse momenta are
factorized so that their propagators are attached to the external
quark lines (see Ref.~\cite{egtns} for detail). Such terms are
absent in Eq.~(\ref{eqfqcd}) because gluon propagators cannot be
attached to the photon lines.  The last term in the both
Eqs.~(\ref{eqfqcd},\ref{eqhqcd}) corresponds to the case when a
$t$ -channel intermediate quark-antiquark pair factorizes
amplitude $T$ into a convolution of two on-shell amplitudes. When
$\alpha_s$ is kept fixed, $A(\omega)$ is replaced by $\alpha_s$
and $D^{(\pm)}_{QCD}$ of Eq.~(\ref{d}) is changed for\footnote{The
sign of Eq.~(31) in Ref.~\cite{egtns} is wrong, however this
misprint does not affect the results of the paper.}
\begin{equation}\label{dfix}
\tilde{D}^{(\pm)}_{QCD}  =
\Big(-\frac{C_F}{2N}\Big)\Big(-\frac{4\alpha^2_s}{\omega^2}\Big)[1
\mp 1]~.
\end{equation}
 The relation $\tilde{D}^{(+)}_{QCD} =0$ means
that DL contributions of the non-ladder Feynman graphs cancel each
other in expressions for the forward scattering amplitudes with
the positive signatures. It was first noticed in Ref.~\cite{gln}
in the QED context and remains true in QCD when $\alpha_s$ is
fixed. According to Eq.~(\ref{d}), accounting for the running
$\alpha_s$ effects violates it. The expression (\ref{dfix}) for
$\tilde{D}^{(-)}_{QCD}$ (as well as Eq.~(\ref{dint}) for
$D^{(\pm)}_{QCD}$) consists of two factors (each in the brackets).
The first factor $(-C_F/2N)$ comes from simplifying the color
structure $t_a t_b t_a t_b$ of the involved graphs $(t_{a,b}$ are
the $SU(3)$-generators) whereas the second factor comes from
integration over momenta of the virtual partons. The terms in
squared brackets in  Eq.~(\ref{d}) correspond to $[\alpha_s(s) \pm
\alpha_s(-s)]$ and the exponential in this equation corresponds to
integration of $\alpha_s(k^2_{\perp})/k^2_{\perp}$. We stress that
the definition of $D_{QCD}$ in Eq.~(\ref{d}) differs from the
definition $D$ in Ref.~\cite{egtns}: $D_{QCD}  = \omega D$.
%The first term in the squared brackets in  corresponds to accounting
%for DL contributions whereas the second term accounts for SL contributions.
Solution to Eq.~(\ref{eqhqcd}) is
\begin{equation}\label{hqcd}
H^{(\pm)}_{QCD} = \frac{\omega - \sqrt{\omega^2 -
B^{(\pm)}_{QCD}}}{2}~,
\end{equation}
with
\begin{equation}\label{bqcd}
B^{(\pm)}_{QCD} = 4b^{(\pm)}_{QCD} = [4\pi A C_F +
D^{(\pm)}]/(2\pi^2) ~.
\end{equation}

 In order to specify the general solution of Eq.~(\ref{eqfqcd}),
 %$K(\omega)$ in Eq.~(\ref{gensolqcd})
 we use (see Ref.~\cite{egtns}) the matching
\begin{equation}\label{matchqcd}
F^{(\pm)}_{QCD}(\omega, y)|_{y = 0} =
\widetilde{F}^{(\pm)}_{QCD}(\omega)~,
\end{equation}
with $\widetilde{F}^{(\pm)}_{QCD}$ corresponding to the DIS off a
nearly on-shell photon (with $Q^2 = \mu^2$). It obeys the new IREE
(cf Eq.~(\ref{eqfqcd})):
\begin{equation}\label{eqfqcdtilde}
\omega \widetilde{F}^{(\pm)}_{QCD} = e^2_q \delta q(\omega) +
H^{(\pm)}_{QCD} \widetilde{F}^{(\pm)}_{QCD}
\end{equation}
where $e_q$ is the electric charge of the initial quark and
$\delta q(\omega)$ is the initial quark density in the $\omega$
-space. In contrast to Eq.~(\ref{eqfqcd}), there is the Born
contribution in the rhs of Eq.~(\ref{eqfqcdtilde}) because in this
case we keep $Q^2 \sim \mu^2$, so the Born term depends on $\mu$
and does not vanish when differentiated with respect to $\mu$.

Eventually we arrive at the final answer for the
 non-singlet structure functions $f^{(\pm)}_{QCD}$ in QCD:
\begin{equation}\label{fnsqcd}
f^{(\pm)}_{QCD} = \frac{e^2_Q}{2}\int_{-\imath \infty }^{\imath
\infty}\frac{d\omega}{2\pi \imath} \big(1/x\big)^{\omega}
\frac{\omega}{\omega - H^{(\pm)}_{QCD}}\, \delta q\, e^{y
H^{(\pm)}_{QCD}} ~.
\end{equation}
Although Eq.~(\ref{fnsqcd}) is obtained for $Q^2 \gg \mu^2$, the
shift $Q^2 \to Q^2+ \mu^2$ generalizes Eq.~(\ref{fnsqcd}) to the
small-$Q^2$ region (see Ref.~\cite{egtsmallq} for detail).  The
small-$x$ asymptotics of $f^{(\pm)}_{QCD}$ is
\begin{equation}\label{asfns}
f^{(\pm)}_{QCD}  \sim (1/x)^{\Delta^{(\pm)}_{QCD}}
\end{equation}
where $\Delta^{(\pm)}_{QCD}$ are called the intercepts.
Straightforwardly they can be found with applying the saddle-point
method to Eq.~(\ref{fnsqcd}). The shorter way is to solve the
equation
\begin{equation}\label{intqcd}
\omega^2- B^{(\pm)}_{QCD} = 0
\end{equation}
for the leading singularity position and to choose its largest
root. The root corresponds to the rightmost singularity of
Eq.~(\ref{fnsqcd}). Ref.~\cite{egtns} reads that
$\Delta^{(+)}_{QCD}= 0.39$ and $\Delta^{(-)}_{QCD} = 0.42$.

\section{Electromagnetic DL corrections to the  non-singlet structure functions}

As exchanges of virtual gluons cannot be isolated from the virtual
photon exchanges, it is necessary to add the electromagnetic (EM)
DL contributions to the QCD expression of Eq.~(\ref{fnsqcd}) for
the non-singlet structure functions. Generalization of
Eq.~(\ref{eqfqcd}) for amplitudes $T^{(\pm)}$ to account for
exchanges of virtual gluons and photons can be done in a very
simple way: with replacing $H^{(\pm)}_{QCD}$ by new non-singlet
anomalous dimensions $h^{(\pm)}_{EM}$ accounting for both EM and
QCD DL contributions.  The IREE for $h^{(\pm)}_{EM}$ is  similar
to Eq.~(\ref{eqhqcd}):

\begin{equation}\label{eqhem}
\omega h^{(\pm)}_{EM}(\omega) = \frac{b_{EM}}{8 \pi^2}  +
(h^{(\pm)}_{EM}(\omega))^2~.
\end{equation}
It changes Eq.~(\ref{fnsqcd}) for a quite similar
 expression
\begin{equation}\label{fnsem}
f^{(\pm)}_{EM} = \frac{e^2_q}{2}\int_{-\imath \infty }^{\imath
\infty}\frac{d\omega}{2\pi \imath} \big(1/x\big)^{\omega}
\frac{\omega}{\omega - H^{(\pm)}_{EM}}\, \delta q\, e^{y
H^{(\pm)}_{EM}}
\end{equation}
where new anomalous dimension $H^{(\pm)}_{EM}$ sums the both QCD
and EM double logarithms. It also looks like $H^{(\pm)}_{QCD}$:
\begin{equation}\label{hem}
H^{\pm}_{EM} = \frac{\omega - \sqrt{\omega^2 -
B^{(\pm)}_{EM}}}{2}~,
\end{equation}
Similarly to  Eq.~(\ref{bqcd}), $B^{(\pm)}_{EM}$ is expressed
through $b^{(\pm)}_{EM}$:
\begin{equation}\label{bpmem}
B^{(\pm)}_{EM}  = b^{(\pm)}_{EM}/(2\pi^2)~.
\end{equation}
 Now let us specify  $b^{(\pm)}_{EM}$:
\begin{equation}\label{bem}
b^{(\pm)}_{EM}  = b^{(\pm)}_{QCD} + a_{\gamma} + D^{(\pm)}_{EM}
\end{equation}
where $a_{\gamma}$  is the electric charge of the quark:
\begin{equation}\label{aem}
a_{\gamma} = e^2_q = 4 \pi \alpha Q^2_q
\end{equation}
and
\begin{equation}\label{dem}
D^{(\pm)}_{EM} =D^{(\pm)}_{g\gamma} + D^{(\pm)}_{\gamma g} +
D^{(\pm)}_{\gamma\gamma}~,
\end{equation}
with
\begin{eqnarray}\label{dgamma}
&&D^{(\pm)}_{g\gamma} = - \frac{4 \alpha Q^2_qC_F}{b}[1\mp
1]e^{\omega\eta}
 \int_{-1}^{\infty} dt e^{-\omega\eta t} \ln t~,
\qquad D^{(\pm)}_{\gamma\gamma} =
-\frac{4\alpha^2 Q^4_q  [1 \mp 1]}{\omega^2}~,\\
\nonumber &&D^{(\pm)}_{\gamma g}  = -\frac{4\alpha Q^2_q C_F}{ b}
\int_0^{\infty} d \rho e^{-\omega \rho}\Big[
\frac{\rho(\rho+\eta)}{(\rho + \eta)^2 + \pi^2} \mp
\frac{\rho}{\rho + \eta}\Big]~.
\end{eqnarray}
When $\alpha_s$ is fixed, the expressions for
$D^{(\pm)}_{g\gamma}$ and $D^{(\pm)}_{\gamma g}$ become simpler:
\begin{eqnarray}\label{dgammafix}
D^{(\pm)}_{g\gamma} =D^{(\pm)}_{\gamma g} =-\frac{4\alpha Q^2_q
\alpha_s C_F}{\omega^2}[1 \mp 1]~.
\end{eqnarray}
Let us explain how $D^{(\pm)}_{ik}$ of Eq.~(\ref{dgamma}) can be
obtained from the QCD expressions for $D^{(\pm)}_{QCD}$ in
Eq.~(\ref{dint}-\ref{dfix}). Eq.~(\ref{dint}) reads  that
$D^{(\pm)}_{QCD}$ contains the QCD couplings depending on
different arguments.

\textbf{(a)}: There is $\alpha(k^2_{\perp})$ that comes when the
soft virtual  gluon with momentum $k^2 \approx -k^2_{\perp}$
 is coupled to quarks.

 \textbf{(b):} There is the sum $[\alpha_s(s) \mp \alpha_s(-s)]$
 from the hard virtual gluon coupled to the quarks.
 In $D^{(\pm)}_{\gamma g}$ and $D^{(\pm)}_{g \gamma}$ one of the
 gluons is replaced by the photon with the same momentum. In contrast  to $\alpha_s$,
 we treat $\alpha$ as fixed: $\alpha= 1/137$~.

 Therefore, when the soft gluon is replaced by the soft photon,
 $\alpha(k^2_{\perp})$ in Eq.~(\ref{dint}) should be replaced by $\alpha
 Q^2_q$ and we arrive at $D^{(\pm)}_{\gamma g}$~.
 Instead, when the hard gluon is replaced, $[\alpha_s(s) \mp
 \alpha_s(-s)]$ should be replaced by $\alpha Q^2_q[1 \mp 1]$,  the remaining
 integration over $k^2_{\perp}$ can easily be done  and we obtain $D^{(\pm)}_{g \gamma}$.
 At  last, combining both previous cases leads us to $D^{(\pm)}_{\gamma
 \gamma}$ where the both gluons are replaced by photons. This case is similar to
 Eq.~(\ref{dfix}), save the color factor  $-C_F/(2N)$. Obviously, the
 replacements the gluons by photons change the  two-gluon color factor $t_at_bt_at_b =
 -C_F/(2N)$ for either $t_at_a = C_F$ (for $D^{(\pm)}_{\gamma g}$ and
 $D^{(\pm)}_{g \gamma}$) or $1$ (for $D^{(\pm)}_{\gamma \gamma}$).

In the QCD framework, the only difference between the small-$x$
behavior of $f^{(\pm)}_u$ (for up-quarks) and $f^{(\pm)}_d$ (for
down-quarks) is the difference between the initial quark densities
$\delta u$ and $\delta d$ whereas both the $x$ and $Q^2$
-evolutions of the initial up- ($u$) and down- ($d$) quark are
identical,  so the subscripts $u$ and $d$ at $f^{(\pm)}_{u,d}$ are
often dropped. Accounting for EM contributions brings a difference
of the both evolutions on the flavor. To mark this difference, we
introduce the non-singlet structure functions, $f^{(\pm)}_u$ and
$f^{(\pm)}_d$, with the subscripts showing the flavor of the
initial quark. Obviously, $f^{(\pm)}_u \neq f^{(\pm)}_d$ even if
$\delta u = \delta d$. As could be well-expected,
Eq.~(\ref{fnsem}) shows that the impact of EM correction on the
small-$x$ behavior of $f^{(\pm)}$ is very small. Indeed, the
estimate of the  impact $\epsilon_{EM}$ of the EM corrections on
the
 intercepts is:
\begin{equation}\label{emimpact}
\epsilon^{(+)}_{EM} = \frac{\Delta^{(+)}_{EM}
 - \Delta^{(+)}_{QCD}}{\Delta^{(+)}_{QCD}} \approx
 \epsilon^{(-)}_{EM} = \frac{\Delta^{(-)}_{EM}
 - \Delta^{(-)}_{QCD}}{\Delta^{(-)}_{QCD}} \approx 1 \%.
\end{equation}

\section{Inclusion of electroweak  DL contributions}
In order to include into consideration all electroweak DL
contributions, adding to the gluon  and photon exchanges, the $W$
and $Z$ -exchanges, we should modify the method that we used in
the previous Sects. by the following reasons:

\textbf{(i)} As the gauge group of the electroweak  interactions
is broken and electroweak bosons become massless photons and
massive $W,Z$  -bosons, the non-singlet structure functions
acquire dependence on the both $\mu$  and $M_{W,Z}$.

\textbf{(ii)} $W$-exchanges cause mixing of $u$ and $d$ -quarks,
so IREE for $f^{(\pm)}_u$ and $f^{(\pm)}_d$ together with IREE for
the anomalous dimensions, are not separable (as in QCD).

Before composing the IREE, let us introduce necessary notations.
We use the notation $g_{W}$ for the $W$-coupling to quarks.  It
does  not depend on the quark flavor.  On the contrary, both the
photon coupling $e_q$ and the $Z$ -boson coupling $g_{qZ}$ to
quarks are flavor-dependent. All these coupling are expressed
through the $SU(3)$ Standard Model coupling $g$ and the Weinberg
angle $\theta$:
\begin{eqnarray}\label{gwz}
 &&g_{uW} = g_{dW} \equiv g_W =  g /\sqrt{2}~,\qquad e_q = g \sin \theta_W
 Q_q = g \sin \theta_W (T_3 + Y/2)~,\\ \nonumber
 &&g_{qZ} = (g/\cos \theta_W)(T_3 - Q_q \sin^2
 \theta_W) = (g/\cos \theta_W)(\cos^2 \theta_W T_3 -  \sin^2
 \theta_W(Y/2))~.
\end{eqnarray}
We keep through  the paper the standard notations $T_3, Y$ and $Q$
for the isospin, hypercharge and electric charge of quarks
together with the standard relation $Q = T_3 + Y/2$.
 We simplify the $M_{W,Z}$ -dependence of the non-singlets, assuming
 that  in the  logarithmic expressions
\begin{equation}\label{masses}
M_W \approx M_Z = M~.
\end{equation}
 Again, it is convenient to introduce the Compton
amplitudes $T^{(\pm)}_u,~T^{(\pm)}_d$
 related to the non-singlet
structure functions  by Eq.~(\ref{compt}). We will address them as
the forward Compton amplitudes, although at energies $\sqrt{s} \gg
M_{W,Z}$ and $Q^2 \gtrsim M^2_{W,Z}$ the lepton and hadron
participating in the DIS can exchange with $\gamma, Z$ (neutral
lepton currents) and $W$ (charged lepton currents). In order to
avoid overloading the  paper we consider only the case of small
$Q^2$:
\begin{equation}\label{smq}
Q^2 \ll M^2_{W,Z}
\end{equation}
where the photon exchange between the lepton and quarks prevails.
The other cases can be considered quite similarly. Under the
approximation of Eq.~(\ref{masses}), the non-singlet functions
$f^{(\pm)}_{u,d}$ and the Compton amplitudes $T^{(\pm)}_{u,d}$
  depend on $s, Q^2$ and the mass scales $\mu$ and $M$.
We assume the following relations between the parameters $s,
~Q^2,~M^2~,\mu^2$~:

\begin{equation}\label{relsqmmu}
s \gg M^2  \gtrsim Q^2 \gg \mu^2~.
\end{equation}
It is  convenient to introduce the amplitudes
$F^{(\pm)}_{u,d}(\omega, y, z)$ related to amplitude
$T^{(\pm)}_{u,d}$ similarly to Eq.~(\ref{ft}):
\begin{equation}
\label{mellinew} T^{(\pm)}_{u,d} = \int_{-\imath \infty }^{\imath
\infty}\frac{d\omega}{2\pi \imath}
\Big(\frac{s}{\mu^2}\Big)^{\omega} \xi^{(\pm)}(\omega)
F_{u,d}^{(\pm)}(\omega, y, z)
\end{equation}
where new variable $z$ is introduced: $z = \ln(M^2/\mu^2)$~.
 In accounting for DL contributions, $\mu$ acts as an infrared
cut-off for DL terms involving soft gluons and photons whereas $M$
 acts as the second cut-off when DL terms involving soft $W,Z$ -bosons
are considered. In contrast to the considered above QCD and EM
cases, IREE for $F^{(\pm)}_{u,d}(\omega, y, z)$ involve the matrix
of new anomalous dimensions $h^{(\pm)}_{ik}$, with $i,k$ being $=
u,d$, and involve the derivatives with respect to $y$ and $z$~:
\begin{eqnarray}\label{eqfew}
&&(\omega + \partial/ \partial y + \partial/ \partial z)
F^{(\pm)}_u = h^{(\pm)}_{uu}(\omega,z) F^{(\pm)}_u +
h^{(\pm)}_{ud}(\omega,z) F^{(\pm)}_d, \\ \nonumber &&(\omega
+\partial/\partial y + \partial/ \partial z) F^{(\pm)}_d =
h^{(\pm)}_{du}(\omega,z) F^{(\pm)}_u + h^{(\pm)}_{dd}(\omega,z)
F_d^{(\pm)}~.
\end{eqnarray}
The anomalous dimensions  $h_{ik}^{(\pm)}$ should be calculate
independently. After they have been found, it is  possible to find
general  solutions to Eqs.~(\ref{eqfew}). In order to specify
them, we will use the matching
\begin{equation}\label{matchunb}
F^{(\pm)}_{u,d}(\omega,y,z)|_{y = 0} =
\widetilde{F}^{(\pm)}_{u,d}(\omega,z)
\end{equation}
with the amplitudes $\widetilde{F}^{(\pm)}_{u,d}(\omega, z)$. They
describe the forward Compton scattering, with the EW DL
corrections accounted  for, in the case when the external photon
has the virtuality $\sim \mu^2$, i.e. almost on-shell.
$\widetilde{F}^{(\pm)}_{u,d}$ should be found independently (cf
Eq.~(\ref{matchqcd})). So, before solving  Eqs.~(\ref{eqfew}) we
should find $h^{(\pm)}_{ik}$  and $\widetilde{F}^{(\pm)}_{u,d}$.
 On this step we are going to simplify our
notations. Trough the paper we keep the DL accuracy. It gives us
the right to neglect terms mixing amplitudes with different
signatures. Therefore, all IREE we compose are separable in the
signatures (see Eqs.~(\ref{eqfqcd},\ref{eqhqcd}) and
Eqs.~(\ref{eqfew},\ref{eqhunb})). So, in what follows we basically
drop the signature superscripts $"(\pm)"$ but restore them when it
is necessary.

\section{Electroweak  anomalous dimensions $h_{ik}$}
In the first place let us  focus on obtaining explicit expressions
for $h_{ik}$. We will do  it with obtaining and  solving
appropriate  IREE.

\subsection{IREE for the anomalous dimensions
$h_{ik}$}

 In our approach, in contrast to DGLAP, the
anomalous dimensions can be found with composing and solving
appropriate IREE for them. Equations for $h_{ik}$ can be obtained
as a generalization of Eq.~(\ref{eqhem}):

\begin{eqnarray}\label{eqhik}
&&\big( \omega + \partial /\partial z \big) h_{uu} =
b^{EM}_{uu}/(8\pi^2) + h^2_{uu} + h_{ud} h_{du}~,\qquad
\big(\omega + \partial / \partial z \big) h_{ud} =
b^{EM}_{ud}/(8\pi^2) + h_{uu}h_{ud} + h_{ud} h_{dd}~,\nonumber\\
&&\big( \omega + \partial / \partial z \big) h_{du} =
b^{EM}_{du}/(8\pi^2) + h_{du}h_{uu} + h_{du} h_{dd}~,\qquad
\big(\omega + \partial /\partial z \big) h_{dd} =
b^{EM}_{dd}/(8\pi^2) + h^2_{dd} + h_{ud} h_{du}~.
\end{eqnarray}
The electromagnetic terms $b^{EM}_{uu}$ and $b^{EM}_{dd}$ in
Eq.~(\ref{eqhik}) are actually defined in Eq.~(\ref{bem}):
\begin{equation}\label{bemik}
b^{EM}_{uu} = b_{QCD} + a_{uu}^{EM} + D^{EM}_{uu}~,\qquad
b^{EM}_{dd} = b_{QCD} + a_{dd}^{EM} + D^{EM}_{dd}~,\qquad
b^{EM}_{ud} = b^{EM}_{du}= 0
\end{equation}
where
\begin{equation}\label{aemik}
a_{uu}^{EM}  = 4 \pi \alpha Q^2_u~,\qquad a_{dd}^{EM}  = 4 \pi
\alpha Q^2_d
\end{equation}
and  $D_{uu}^{EM}, D_{dd}^{EM}$ can similarly be taken from
Eqs.~(\ref{dem},\ref{dgamma}), replacing $Q_q$ by $Q_u$ and $Q_d$
respectively. We remind that we have dropped the signature
superscripts $"\pm"$ for the sake of simplicity.  The fact that
$b_{du}^{EM} = _{ud}^{EM} = 0$ simplifies the system in
Eq.~(\ref{eqhik}). It is convenient to re-write Eq.~(\ref{eqhik})
in terms of symmetrized combinations $h_{S,A}$ and  $b^{EM}_{S,A}$
defined as follows:
\begin{equation}\label{hpm}
h_S = h_{uu} + h_{dd}~,\qquad h_A = h_{uu} - h_{dd}~,\qquad
b^{EM}_{S} = b^{EM}_{uu} + b^{EM}_{dd}~,\qquad b^{EM}_{A} =
b^{EM}_{uu} - b^{EM}_{dd}~,
\end{equation}
and to introduce $h$~:
\begin{equation}\label{hhs}
h = - \omega + h_S~.
\end{equation}
 In these terms  Eq.~(\ref{eqhik}) takes the
simpler form:

\begin{eqnarray}\label{eqhiksimple}
&&\frac{\partial h}{\partial z} = b^{EM}_S/(8\pi^2) + \frac{1}{2}
h^2 + \frac{1}{2}h^2_{A} - \frac{\omega^2}{2} + 2h_{ud} h_{du}~,
\\ \nonumber &&\frac{\partial h_{A}}{\partial z} =
b^{EM}_{A}/(8\pi^2) + h_{A}h~,\qquad \frac{\partial
h_{ud}}{\partial z} = h_{ud}h~,\qquad\frac{\partial
h_{du}}{\partial z} = h_{ud}h~.
\end{eqnarray}
Eq.~(\ref{eqhiksimple}) reads that $ h_{ud} = h_{du}$~.

\subsection{General expressions for $h_{ik}$ }

Eqs.~(\ref{eqhik},\ref{eqhiksimple}) for $h_{ik}$ are
non-linear,so solving them exactly is a quite serious technical
problem. We do not pursue this aim in the present paper. Instead,
we suggest an approximative procedure based on the obvious fact
that the QCD coupling is greater than the electroweak ones.  It
means that in Eqs.~(\ref{eqhik},\ref{eqhiksimple})
\begin{equation}\label{bornhik}
b^{EM}_{S} \gg b^{EM}_{A}~, b_{ud}~, b_{du}~.
\end{equation}
Then, Eq.~(\ref{bornhik}) allows to conclude that
\begin{equation}\label{relhik}
h_{S} \gg h_{A}~, h_{ud}~, h_{du}~.
\end{equation}
Using this relation, we can neglect $h^2_{A}$ and $h_{ud} h_{du}$
compared to $h^2_{S}$ in the rhs of the first of equations
Eqs.~(\ref{eqhiksimple}) and write an approximation for
Eqs.~(\ref{eqhiksimple}) :
\begin{eqnarray}\label{eqhapprox}
&&\frac{\partial h}{\partial z} = \frac{b^{EM}_{S}}{8 \pi^2} -
\frac{\omega^2}{2} + \frac{1}{2} h^2~,\qquad\frac{\partial
h_{A}}{\partial z} = \frac{b^{EM}_{A}}{8 \pi^2} + h_{A} h~,
\\ \nonumber  &&\frac{\partial h_{ud}}{\partial z}=
h_{ud} h~,\qquad\qquad\qquad\frac{\partial h_{du}}{\partial z}=
h_{ud} h~.
\end{eqnarray}
The first of Eqs.~(\ref{eqhapprox}) is the Riccatti equation and
the others are linear, so they can be easily solved. The general
solution for $h_{S}$ can be written as
\begin{eqnarray}\label{hikgen}
h_{S}(\omega, z) = \omega + \lambda \frac{1 + C_{S}e^{\lambda
z}}{1 - C_{S}e^{\lambda z}}~,\qquad h_{ud} = h_{du} = C_{ud} \exp
\int^z_0 d t h(\omega,t)~,  \\ \nonumber h_{A} =
\Big[\frac{b^{EM}_{A}}{8 \pi^2} \int^z_0 d t \exp \Big(- \int^t_0
d t' h(\omega,t') \Big) + C_{A} \Big]\exp \int^z_0 d t
h(\omega,t)~,
\end{eqnarray}
with $\lambda = \sqrt{\omega^2 - 2 b^{EM}_{S}/(8\pi^2)}$~. $C_{S},
C_{A}(\omega)$ and $C_{ud}(\omega)$ being an arbitrary functions
of $\omega$. They have to be specified. We do it, invoking the
matching
\begin{equation}\label{matchhunb}
h_{ik}(\omega, z)|_{z = 0} = H_{ik}(\omega)
\end{equation}
where $H_{ik}(\omega)$ are the auxiliary anomalous dimensions
corresponding to the case of the unbroken electroweak symmetry so
that $W,Z$ -bosons are massless, and the cut-off $\mu$ is applied
to all virtual bosons.Combining Eqs.~(\ref{hikgen}) and
(\ref{matchhunb}), we express the unknown functions $C_{S,A,ud}$
in terms of $H_{ik}$~:
\begin{equation}\label{cikh}
C_{S} = - (\lambda - H)/(\lambda + H)~,\qquad C_{A} =H_{A}~,\qquad
C_{ud}= H_{ud}
\end{equation}
where similarly to Eqs.~(\ref{hpm}, \ref{hhs}) we have denoted $H
= - \omega + H_S$ and $H_S = H_{uu} + H_{dd}$~, $H_A = H_{uu} -
H_{dd}$~.

\subsection{Anomalous dimensions at the unbroken EW gauge symmetry}

IREE for $H_{ik}$ differ from Eqs.~(\ref{eqhik})  only in
inhomogeneous terms:
\begin{eqnarray}\label{eqhunb}
&&\omega  H^{(\pm)}_{uu} = b^{(\pm)}_{uu}/(8\pi^2) +
(H^{(\pm)}_{uu})^2 + H^{(\pm)}_{ud} H^{(\pm)}_{du}~,\qquad \omega
 H^{(\pm)}_{ud} = b^{(\pm)}_{ud}/(8\pi^2) + H^{(\pm)}_{uu}H^{(\pm)}_{ud} +
 H^{(\pm)}_{ud} H^{(\pm)}_{dd}~, \\
\nonumber && \omega H^{(\pm)}_{du} = b^{(\pm)}_{du}/(8\pi^2) +
H^{(\pm)}_{du}H^{(\pm)}_{uu} + H^{(\pm)}_{du}
H^{(\pm)}_{dd}~,\quad \omega H^{(\pm)}_{dd} =
b^{(\pm)}_{dd}/(8\pi^2) + (H^{(\pm)}_{dd})^2 + H^{(\pm)}_{ud}
H^{(\pm)}_{du}
\end{eqnarray}
where $b^{(\pm)}_{ik}$ generalize  $b^{EM}$ to the case of the
massless EW bosons. Similarly to Eq.~(\ref{bem}) they can be
represented  as  the sum
\begin{equation}\label{badunb}
b^{(\pm)}_{ik} = \delta_{ik}~b^{(\pm)}_{QCD}   + a_{ik}  +
D^{(\pm)}_{ik}.
\end{equation}
Term $b^{(\pm)}_{QCD}$ in Eq.~(\ref{badunb}) is defined in
Eq.~(\ref{bqcd}), $a_{ik}$  can easily be obtained from
Eq.~(\ref{aem}), adding to $a^{EM}$ the $Z$ and $W$ -boson
couplings:
\begin{equation}\label{aunb}
a_{uu} = a_{dd} = 4\pi \alpha Q^2_u  + g^2_{uZ}  =4\pi
\frac{\alpha}{\sin^2\theta_W} \frac{(1 + Y^2 \tan^2
\theta_W)}{4}~,\qquad a_{ud} = a_{du}  = \frac{g^2}{2}=
\frac{4\pi\alpha}{2\sin^2\theta_W}
\end{equation}
and $D^{(\pm)}_{ik}$ are generalizations  of  $D^{(\pm)}_{EM}$
defined in  Eq.~(\ref{dem}).  It is convenient to represent
$D^{(\pm)}_{ik}$ in the following way (cf Eq.~(\ref{dem})):
\begin{eqnarray}\label{dunb}
D^{(\pm)}_{uu} = D^{(\pm)}_{dd}  =&& -\frac{4 \alpha C_F}{b \sin^2
\theta_W}\Big[\frac{(1+ Y^2\tan^2 \theta_W)}{4} [1\mp 1]
e^{\omega\eta}
 \int_{-1}^{\infty} dt e^{-\omega\eta t} \ln t \; + \\ \nonumber
 &&\int_0^{\infty} d \rho e^{-\omega \rho}
 \Big(\frac{(3+ Y^2\tan^2 \theta_W)}{4} \frac{\rho(\rho+\eta)}{(\rho + \eta)^2 + \pi^2}\,
 \mp \, \frac{(1+ Y^2\tan^2 \theta_W)}{4} \frac{\rho}{\rho + \eta}
\Big)\Big]\; - \\ \nonumber &&\frac{4 \alpha^2}{\omega^2 \sin^4
\theta_W}\Big[[1 \mp 1] \frac{(1+ Y^2\tan^2
\theta_W)^2}{16}\, +\, \frac{(-1+ Y^2\tan^2 \theta_W)}{8}\Big], \\
\nonumber D^{(\pm)}_{ud} = D^{(\pm)}_{du} =&& - \frac{2\alpha
C_F}{b \sin^2 \theta_W}[1 \mp 1] e^{\omega\eta}
 \int_{-1}^{\infty} dt e^{-\omega\eta t} \ln t \;
 \pm \frac{2 \alpha C_F}{b \sin^2
\theta_W}\int_0^{\infty} d \rho e^{-\omega \rho} \frac{\rho}{\rho
+ \eta}\; - \\ \nonumber &&\frac{4\alpha^2}{\omega^2 \sin^4
\theta_W} [1 \mp 2] \Big[ \frac{(-1+ Y^2\tan^2 \theta_W)}{8}\Big].
\end{eqnarray}
When $\alpha_s$ is fixed, the expressions for $D^{(\pm)}_{uu}$ and
$D^{(\pm)}_{dd}$ look more simple and instead of Eq.~(\ref{dunb}))
we obtain:
\begin{eqnarray}\label{dunbfix}
D^{(\pm)}_{uu} = D^{(\pm)}_{dd}  = &&- \frac{8 \alpha
\alpha_sC_F}{\omega^2  \sin^2 \theta_W}\frac{(3+ Y^2\tan^2
\theta_W)}{4} [1 \mp 1] - \\ \nonumber &&\frac{4
\alpha^2}{\omega^2 \sin^4 \theta_W} \frac{(1+ Y^2\tan^2
\theta_W)^2}{16}[1 \mp 1] -\frac{4 \alpha^2}{\omega^2 \sin^4
\theta_W} \frac{(-1+ Y^2\tan^2 \theta_W)}{8}~, \\ \nonumber
D^{(\pm)}_{ud} = D^{(\pm)}_{du}= &&- \frac{2\alpha \alpha_s
C_F}{\omega^2 \sin^2 \theta_W}[1 \mp 1] -
\frac{4\alpha^2}{\omega^2 \sin^4 \theta_W} [1 \mp 2] \Big[
\frac{(-1+ Y^2\tan^2 \theta_W)}{8}\Big]~.
\end{eqnarray}
Let us comment on Eqs.~(\ref{dunb},\ref{dunbfix})). The terms
$\sim 1/b$ in Eq.~(\ref{dunb}) (the term $\sim \alpha_s$ in
Eq.~(\ref{dunbfix})) come from interference of the QCD and EW DL
contributions. The next term in the both Eqs. accumulate the DL
contributions of the neutral EW bosons, $\gamma$ and $Z$. A part
of those terms in Eq.~(\ref{dunb}) (all of them in
Eq.~(\ref{dunbfix})) is proportional to the signature factor $[1
\mp 1]$ and therefore vanish when the signature is positive. In
other words, non-ladder DL contributions to the amplitudes with
the positive signature cancel each other totally when couplings
are fixed (and cancel only partly when some of the couplings ar
running)\footnote{We remind that this compensation was first
noticed in Ref.~\cite{gln}  in the QED context.}. The presence of
the last term in Eqs.~(\ref{dunb},\ref{dunbfix})) demonstrates
explicitly that accounting for the $W$-boson exchanges breaks such
a compensation even when the couplings are fixed. Nevertheless, at
fixed $\alpha_s$ summation over flavors in Eq.~(\ref{dunbfix})
leads to the zero contribution of the non-ladder graphs:
\begin{equation}\label{zerod}
D^{(+)}_{uu} + D^{(+)}_{ud} + D^{(+)}_{dd} + {}D^{(+)}_{du} = 0~.
\end{equation}
Eq.~(\ref{zerod}) is quite similar to the QCD result for
$D^{(+)}_{QCD}$ with fixed $\alpha_s$ obtained first in
Ref.~\cite{kl} because summation over flavors in
Eq.~(\ref{dunbfix}) is equivalent to summation over colors in QCD.
As $b_{ik}$ are now fixed, we can solve Eqs.~(\ref{eqhunb}).
Combining Eqs.~(\ref{badunb},\ref{aunb},\ref{dunb}) we see that
$b_{uu}=b_{dd}$~, $b_{ud}=b_{du}$ and therefore Eq.~(\ref{eqhunb})
reads that $H_{uu} = H_{dd}$ and $H_{ud} = H_{du}$~. After that
Eq.~(\ref{eqhunb})  can easily be solved:
\begin{eqnarray}\label{hunb}
&&H_{uu}=H_{dd} = \frac{1}{2}\Big[\omega - E\Big]~,
\\ \nonumber
&&H_{ud}=H_{du}=\frac{\widetilde{b}_{ud}}{E}
\end{eqnarray}
where
\begin{equation}\label{be}
\widetilde{b}_{uu} = \frac{b_{uu}}{8\pi^2}~,\qquad
\widetilde{b}_{ud} = \frac{b_{ud}}{8\pi^2}~,\qquad  E
=\sqrt{\frac{\omega^2- 4 \widetilde{b}_{uu} + \sqrt{(\omega^2 - 4
\widetilde{b}_{uu})^2 - 16 \widetilde{b}^2_{ud}}}{2}}~.
\end{equation}

\subsection{Specifying general expressions for $h_{ik}$}

 Combining Eq.~(\ref{eqhunb}) with Eq.~(\ref{cikh}) and substituting
 them into Eq.~(\ref{hikgen}) leads to explicit expressions  for
 $h_{ik}$~:
\begin{eqnarray}\label{hike}
&&h_{S}(\omega, z) = \omega + \lambda \frac{(\lambda -E) -
(\lambda + E)e^{\lambda z}}{(\lambda - E) + (\lambda +
E)e^{\lambda z}}~,\qquad
h_{ud} = h_{du} = \frac{\widetilde{b}_{ud}}{E} \exp \int^z_0 dt h(\omega,t)~,  \\
\nonumber &&h_{A} = \frac{b^{EM}_{A}}{8 \pi^2}\int^z_0 d t \exp
\Big(- \int^t_0 d t' h(\omega,t') \Big) \exp \int^z_0 d t
h(\omega,t)~.
\end{eqnarray}
Denoting
\begin{equation}\label{beta}
\lambda/E =  \tanh \beta~,
\end{equation}
we obtain that
\begin{equation}\label{hbeta}
h = - \frac{\lambda}{\tanh (\lambda  z/2 + \beta)}~.
\end{equation}
Substituting it into Eq.~(\ref{hike}) leads to explicit
expressions for $h_S, h_A, h_{ud}$~:
\begin{eqnarray}\label{hik}
&&h_S= \omega  -\frac{\lambda}{\tanh (\lambda  z/2 +
\beta)},~~h_{ud} = h_{du} = \frac{\widetilde{b}_{ud}}{E}
\frac{\sinh^2 \beta}{\sinh^2\big(\lambda z/2 + \beta \big)}~, \\
\nonumber &&h_A =\frac{b^{EM}_{A}}{8 \pi^2} \frac{1}{2 \lambda
\sinh^2\big(\lambda z/2 + \beta \big)} \Big[-\lambda z - \sinh
2\beta + \sinh(\lambda z + 2 \beta)\Big]~.
\end{eqnarray}
Eq.~(\ref{hik}) manifests  that the breaking the $SU(3) \otimes
U(1)$ symmetry of the electroweak gauge group leads  to the
non-zero $h_A$ in contrast to the expressions Eq.~(\ref{hunb})
obtained under the assumption of the unbroken EW symmetry.

\section{Auxiliary amplitudes $\widetilde{F}_{u,d}$}

In the present Sect. we calculate the auxiliary amplitudes
$\widetilde{F}_u,~\widetilde{F}_u$ in order to  use them in
Eq.~(\ref{matchunb}). IREE for them are similar to
Eq.~(\ref{eqfew}), save two points: the first is the absence of
the $y$ -dependence because $Q^2 = \mu^2$ for
$\widetilde{F}_{u,d}$ and the second is appearing initial
contributions  because they depend on $\mu$ at $y = 0$:

\begin{eqnarray}\label{eqftilde}
(\omega +  \partial/ \partial z) \widetilde{F}_u
= e^2_u \delta u + h_{uu}(\omega,z) \widetilde{F}_u + h_{ud}(\omega,z) \widetilde{F}_d, \\
\nonumber ~ (\omega +
 \partial/ \partial z) \widetilde{F}_d = e^2_d \delta d + h_{du}(\omega,z) \widetilde{F}_u +
h_{dd}(\omega,z) \widetilde{F}_d~.
\end{eqnarray}
The factors $\delta u, \delta d$ in Eq.~(\ref{eqftilde}) stand for
the initial quark densities in the $\omega$ -space. As the
anomalous dimensions $h_{ik}$ have been found in the previous
Sect. (see Eq.~(\ref{hik})), we can solve Eq.~(\ref{eqftilde}).
Our strategy is to  find a general solution to
Eq.~(\ref{eqftilde}) and after that to specify it with using the
matching to the other auxiliary amplitudes $\phi_{u,d}$ of the
same process, however obtained under the assumption of the
unbroken $SU(2) \otimes U(1)$ symmetry:
\begin{equation}\label{matchfi}
\widetilde{F}_u|_{z=0} = \phi_u,~~~\widetilde{F}_d|_{z=0} =
\phi_d~.
\end{equation}
\subsection{General solution to Eq.~~(\ref{eqftilde})}

Introducing the symmetrized combinations
\begin{equation}\label{ftildesa}
\widetilde{F}_S = \widetilde{F}_u + \widetilde{F}_d,
~~\widetilde{F}_A = \widetilde{F}_u - \widetilde{F}_d,
\end{equation}
we can rewrite Eq.~(\ref{eqftilde}) in the symmetrical form:
\begin{eqnarray}\label{eqftildesa}
\partial \widetilde{F}_S/ \partial z
= (e^2_u \delta u + e^2_d \delta d)+ \Big(-\omega +
\frac{1}{2}h_S(\omega,z)\Big) \widetilde{F}_S +
h_{ud}(\omega,z) \widetilde{F}_S + \frac{1}{2}h_A(\omega,z)\widetilde{F}_A, \\
\nonumber
\partial \widetilde{F}_A/ \partial z
= (e^2_u \delta u -e^2_d \delta d)+ \Big(-\omega +
\frac{1}{2}h_S(\omega,z)\Big) \widetilde{F}_A - h_{ud}(\omega,z)
\widetilde{F}_A + \frac{1}{2}h_A(\omega,z)\widetilde{F}_S .
\end{eqnarray}
It is  easy  to write down a general  solution to
Eq.~(\ref{eqftildesa}) in terms  of integrals of $h_{ik}$.
However, the expressions for $h_{ik}$ are rather complicated,
which makes scarcely possible performing those integrations.
Instead, we obtain an approximative solution to
Eq.~(\ref{eqftildesa}),  having noticed that according to
Eq.~(\ref{hik}) $h_S  \gg h_A, h_{ud}$. It gives us  the right to
drop the term $h_A  \widetilde{F}_A$ in the first of
Eq.~(\ref{eqftildesa}). After that we arrive at the following
results:

\begin{eqnarray}\label{ftildegen}
&&\widetilde{F}_S = \Big[\phi_S(\omega) + c_S(\omega)\int_{0}^{z}
dt e^{- \Psi(\omega, t)}\Big] e^{\Psi(\omega, z)}~, \\ \nonumber
&&\widetilde{F}_A =\Big[\phi_A(\omega) + c_A (\omega)\int_0^z dt
e^{- \Psi(\omega, t)}
  + \frac{\phi_S}{2} \int_{0}^{z} dt h_A (\omega,t) + \frac{c_S}{2}
  \int_{0}^{z} dt h_A (\omega,t) \int_0^t
  d v
e^{- \Psi(\omega, v)}\Big] e^{\Psi(\omega, z)}
\end{eqnarray}
where $c_S =e^2_u \delta u + e^2_d\, \delta d,~ c_A = e^2_u \delta
u -e^2_d\, \delta d$ and
\begin{equation}\label{psi}
\Psi (\omega,z) = \int_0^z dt \Big[- \omega  +
\frac{1}{2}h_S(\omega, t)\Big] = - \frac{\omega z}{2} - \ln
\Big(\frac{\sinh (\lambda z/2 + \beta)}{\sinh \beta}\Big)~.
\end{equation}
Obviously, $\widetilde{F}_S = \phi_S$ and $\widetilde{F}_A =
\phi_A$ at $z = 0$ in accordance with  the matching of
Eq.~(\ref{matchfi}). Now we should find $\phi_{S,A}$ in order to
specify Eq.~(\ref{ftildegen}).

\subsection{Amplitudes $\phi_{u,d}$}

Amplitudes $\phi_{u,d}$ describe  the forward Compton scattering
off $u$ and $d$  -quarks under  the assumption of unbroken EW
symmetry and with the photon being on-shell. Obviously, they obey
the following IREE:

\begin{eqnarray}\label{eqfi}
&&\omega \phi_u = e^2_u \delta u + H_{uu} \phi_u + H_{ud} \phi_d~, \\
\nonumber &&\omega \phi_d = e^2_d\, \delta d + H_{du} \phi_u +
H_{dd} \phi_d~,
\end{eqnarray}
with the obvious solution:
\begin{equation}\label{fi}
\phi_S \equiv \phi_u + \phi_d = \frac{c_S}{\omega - H_{uu} -
H_{ud}}~,\qquad \phi_A \equiv \phi_u - \phi_d = \frac{c_A}{\omega
- H_{uu} + H_{ud}}~.
\end{equation}
 We have
used in Eq.~(\ref{fi}) that $H_{uu} = H_{dd}$~.

\subsection{Specifying the general solutions for $\widetilde{F}_{S,A}$}

When $\phi_A$ and $\phi_S$ are known, the general expressions in
Eq.~(\ref{ftildegen}) can be specified:
\begin{eqnarray}\label{ftilde}
&&\widetilde{F}_S  = c_S \Big[ \frac{e^{-\omega z/2}\sinh
\beta}{(\omega - H_{uu} - H_{ud})\sinh (\lambda z/2 + \beta)}\,
+\, \frac{4 \sinh (\lambda z/4)\cosh (\lambda z/4 + \beta -
\varphi)}{\sqrt{\omega^2 - \lambda^2}\sinh (\lambda z/2 +
\beta)}\Big]~,
\\ \nonumber
&&\widetilde{F}_A =c_A \Big[\frac{e^{-\omega z/2}\sinh
\beta}{(\omega- H_{uu} + H_{ud})\sinh (\lambda z/2 + \beta)}\, +\,
\frac{4 \sinh (\lambda z/4)\cosh(\lambda z/4 + \beta -
\varphi)}{\sqrt{\omega^2 - \lambda^2}\sinh (\lambda z/2 + \beta)}
\Big]\, +\, \frac{c_S}{2} \frac{e^{-\omega z/2}}{\sinh (\lambda z/2 + \beta)}\cdot \\
\nonumber &&\Big[\frac{\sinh \beta}{(\omega- H_{uu} -
H_{ud})}\int_0^z dt\, h_A(\omega, t)\, +\, \frac{4}{\sqrt{\omega^2
-\lambda^2}} \int_0^z dt\, h_A(\omega,t)\,e^{\omega t/2}\sinh
(\lambda t/4)\cosh(\lambda t/4 + \beta - \varphi) \Big]~,
\end{eqnarray}
where we have used the notation
\begin{equation}\label{varphi}
\lambda / \omega = \tanh\varphi~.
\end{equation}

\section{Explicit expressions for the electroweak amplitudes $F_{u,d}$}

In the previous Sects. we obtained explicit expressions for  the
electroweak anomalous dimensions $h_{ik}$ and the auxiliary
amplitudes $\widetilde{F}_{u,d}$~. Therefore, we can now find
solutions to Eq.~(\ref{eqfew}) for amplitudes $F_{u,d}$~. As
Eq.~(\ref{eqfew}) is quite similar to Eq.~(\ref{eqftilde}),
solving it can be done in the same way. Again it is convenient to
introduce the symmetrized notations
\begin{equation}\label{fsafud}
F_S = F_u + F_d~,\qquad F_A = F_u - F_d
\end{equation}
and express the solution in terms  of them. Obviously,
\begin{eqnarray}\label{fsa}
&&F_S(\omega, z-y,z) = \widetilde{F}_S (\omega,
z-y)e^{\Psi(\omega,z) - \Psi(\omega, (z-y))} = \\ \nonumber &&c_S
(\omega) \frac{e^{-\omega z/2}}{\sinh (\lambda z/2 +
\beta)}\Big[\frac{\sinh \beta}{\omega- H_{uu} - H_{ud}}
+\int_0^{z-y} dt e^{\omega t/2}\sinh (\lambda t/2 + \beta)\Big]~,
\\ \nonumber
&&F_A (\omega, z-y,z)= \Big[ \widetilde{F}_A (\omega, z-y) +
\frac{1}{2} \int_{z-y}^z d t h_A(\omega,  t)F_S(\omega,z-y, t)
e^{-\Psi (\omega,t) + \Psi (\omega, z-y)}\Big]e^{\Psi(\omega, z) -
\Psi (\omega, z-y)} = \\ \nonumber &&\frac{e^{-\omega z/2}}{\sinh
(\lambda z/2 + \beta)} \Big[c_A \Big(\frac{\sinh \beta}{\omega -
H_{uu} + H_{ud}} + \int_0^{z-y} dt e^{\omega t/2}\sinh (\lambda
t/2 + \beta) \Big) +
 \frac{c_S}{2} \Big(\frac{\sinh \beta}{\omega - H_{uu}
- H_{ud}} \int_0^z dt h_A (t) \\ \nonumber &&+
 \int_0^{z-y} dt h_A (t)\int_0^t du e^{\omega u/2}\sinh (\lambda u/2 +
\beta) + \int^z_{z-y} dt h_A (t)\int_0^{z-y} du e^{\omega
u/2}\sinh (\lambda u/2 + \beta)\Big)\Big]
\end{eqnarray}
where $\widetilde{F}_S$ and $\widetilde{F}_A$ are defined in
Eq.~(\ref{ftilde}) and $h_A$ is given by Eq.~(\ref{hik}). We
remind that $c_S = e^2_u \delta u + e^2_d\, \delta d$ and $c_A =
e^2_u \delta u - e^2_d\, \delta d$.

\section{Expressions for the non-singlet structure functions}

 Now we can write
down explicit expressions for the non-singlet structure functions
including the total resummation of QCD and EW double-logarithmic
contributions.  We express the non-singlet structure function
$f_u$ of $u$ -quark and the non-singlet structure function $f_d$
of $d$ -quark in terms of their symmetrized combinations $f_S$ and
$f_A$:
\begin{equation}\label{fpmsa}
f_S = f_u + f_d~,\qquad f_A = f_u - f_d~.
\end{equation}
Combining Eqs.~(\ref{ft}) and (\ref{fsa}) leads us to the
following expressions:
\begin{eqnarray}\label{fnsew}
&&f_S = \frac{1}{2} \int_{- \imath \infty}^{\imath \infty} \frac{d
\omega}{2 \pi \imath} \Big(\frac{s}{\mu^2}\Big)^{\omega} F_S
(\omega, z,y)~,
\\ \nonumber &&f_A = \frac{1}{2} \int_{- \imath
\infty}^{\imath \infty} \frac{d \omega}{2 \pi \imath}
\Big(\frac{s}{\mu^2}\Big)^{\omega} F_A (\omega, z,y)~.
\end{eqnarray}
The Mellin amplitudes $F_{S,A}$ in Eq.~(\ref{fnsew}) are given by
Eq.~(\ref{fsa}). When the non-singlet structure functions $f_u,
f_d$ are calculated in the QCD framework, the difference between
them, $f_A  \neq 0$, only if $c_A = e^2_u \delta u - e^2_d\,
\delta d \neq 0$. Including the EW corrections changes the
situation cardinally. Indeed, Eq.~(\ref{fsa}) manifests that the
expression for $f_A$ includes the contribution proportional to
$c_A$ and, in addition, the contribution proportional to $c_S =
e^2_u \delta u + e^2_d\, \delta d$. The latter contribution arises
because of mixing $u$ and $d$ -quarks through $W$ -boson
exchanges. It means  that, with the EW corrections accounted for,
$f_u \neq f_d$ even if $c_A = 0$. We remind that
Eqs.~(\ref{fnsew}) describe $f_u$ and $f_d$ in the region
(\ref{relsqmmu}).

\section{Impact of the EW double-logarithms on the non-singlet intercepts}

Let us consider the small-$x$ asymptotics of $f^{(\pm)}_u$ and
$f^{(\pm)}_d$. When they are calculated in the QCD framework, they
are identical, save the difference between $e^2_u \delta u$ and
$e^2_d \delta d$, and given by Eq.~(\ref{asfns}). Accounting for
the EW DL contributions keeps the Regge form of the asymptotics
but changes the QCD intercepts $\Delta^{(\pm)}_{QCD}$ for the new
ones which we denote $\Delta^{(\pm)}$. According to
Eq.~(\ref{intqcd}), the intercepts are the rightmost singularities
of $F_{S,A}$ in Eq.~(\ref{fnsew}) The leading singularity is the
square root branching point in Eq.~(\ref{be}):
\begin{equation}\label{branch}
(\omega^2 - 4b_{uu}/(8 \pi^2))^2 -16 (b_{ud}/(8 \pi^2))^2=0~.
\end{equation}
The terms $b_{uu},~b_{ud}$ in Eq.~(\ref{branch}) are defined in
Eq.~(\ref{badunb}). They depend on the signatures, so from now on
we should once more write explicitly the signature superscripts
$"\pm"$. It is interesting to note that Eq.~(\ref{branch})
corresponds to the unbroken $SU(3)\otimes SU(2)\otimes U(1)$ gauge
symmetry and therefore can be rewritten in the following way:
\begin{equation}\label{eqintc}
\omega^2 =
%\frac{a_{uu} + a_{ud} + D^{(\pm)}_{uu} +
%D^{(\pm)}_{ud}}{2 \pi^2} =
\frac{2}{\pi} \Big[A(\omega) C_F + \alpha_{SU(2)}C'_F +
\alpha_{U(1)}(Y/2)^2 \Big] + \frac{D^{(\pm)}}{2 \pi^2}
\end{equation}
where  $\alpha_{SU(2)}  = \alpha/\sin^2 \theta_W ,~\alpha_{U(1)} =
\alpha/\cos^2 \theta_W$; then, $C'_F = 3/4,~N' = 2,~Y = 1/3$ and
\begin{eqnarray}\label{dtot}
&&D^{(\pm)} = D_{QCD}^{(\pm)} + \zeta \frac{2 \alpha^2_{SU(2)}
C'_F}{\omega^2 N'} - z \frac{4 \alpha^2_{U(1)}Y^4}{16\omega^2} -
\frac{4 \alpha_{SU(2)} C_F C'_F}{b} \Big[\int_0^{\infty} d \rho
e^{-\omega \rho}\Big(\frac{\rho (\rho + \eta)}{(\rho + \eta)^2 +
\pi^2} \\
\nonumber  &&\mp \frac{\rho}{\rho + \eta}\Big) +  \zeta
e^{\omega\eta}
 \int_{-1}^{\infty} dt e^{-\omega\eta t} \ln t\Big]
 -\frac{4 \alpha_{U(1)} C_F Y^2}{4 b} \Big[\int_0^{\infty} d \rho
 e^{-\omega \rho} \Big(\frac{\rho (\rho +
\eta)}{(\rho + \eta)^2 + \pi^2}  \mp \frac{\rho}{\rho + \eta}\Big) + \\
\nonumber &&\zeta e^{\omega\eta}
 \int_{-1}^{\infty} dt e^{-\omega\eta t} \ln t\Big] - \zeta \frac{8
 \alpha_{SU(2)}
 \alpha_{U(1)}C'_F Y^2}{4 \omega^2}~.
\end{eqnarray}
In Eq.~(\ref{dtot}) we have denoted $\zeta = [1 \mp 1]$~.

When $\alpha_s$ is assumed fixed, Eq.~(\ref{eqintc}) looks more
simple:
\begin{equation}\label{eqintcfix}
\omega^2 - a - d^{(\pm)}/\omega^2 = 0~,
\end{equation}
with  \begin{eqnarray} \label{adtot} &&a = \frac{8\alpha_s}{3\pi}
+ \frac{3 \alpha}{2\pi \sin^2 \theta_W} + \frac{\alpha}{18 \pi
\cos^2 \theta_W}~,\qquad d^{(+)} = 0~, \\ \nonumber &&d^{(-)} =
\frac{1}{2\pi^2}\Big[\frac{8}{9} \alpha^2_s - 8\frac{\alpha_s
\alpha}{\sin^2 \theta_W} - \frac{8}{27}\frac{\alpha_s
\alpha}{\cos^2 \theta_W}  + \frac{3}{4}\frac{\alpha^2}{\sin^4
\theta_W} -\frac{1}{6}\frac{\alpha^2}{\sin^2 \theta_W \cos^2
\theta_W} - \frac{1}{324}\frac{\alpha^2}{\cos^4 \theta_W}\Big]~.
\end{eqnarray}
Eq.~~(\ref{eqintcfix}) can easily be solved analytically, the
solutions, $\omega^{(\pm)}_0$ are
\begin{equation}\label{intfix}
\omega^{(+)}_0 = \sqrt{a}~,\qquad \omega^{(-)}_0= \sqrt{(a +
\sqrt{a^2 + 4 d^{(-)}})/2}~.
\end{equation}
On the contrary, Eq.~(\ref{eqintc}) cannot be solved analytically.
Numerical solutions to Eq.~(\ref{eqintc}) depend on $\eta$ and
their maximums which we call the intercepts\footnote{See
Ref.~\cite{egtns,egtsmallq} for detail.} are

\begin{equation}\label{deltapm}
\Delta^{(+)} = 0.373~,\qquad \Delta^{(-)} = 0.354~,
\end{equation}
while the QCD intercepts $\Delta^{(\pm)}_{QCD}$ obtained in
Ref.~\cite{egtns} are
\begin{equation}\label{deltaqcd}
\Delta^{(+)}_{QCD} = 0.385~,\qquad \Delta^{(-)}_{QCD} = 0.423~.
\end{equation}
However, the QCD intercepts of Eq.~(\ref{deltaqcd}) include both
DL and single-logarithmic (SL) contributions. When, in addition to
DL terms, only the SL terms contributing to $\alpha_s$ are taken
into account and other SL terms are neglected, the QCD non-singlet
intercepts $\widetilde{\Delta}^{(\pm)}_{QCD}$  differ from
$\Delta^{(\pm)}_{QCD}$~:
\begin{equation}\label{deltaqcddl}
\widetilde{\Delta}^{(+)}_{QCD} = 0.346~,\qquad
\widetilde{\Delta}^{(-)}_{QCD} = 0.389~.
\end{equation}
Therefore, the impacts $\epsilon^{(\pm)}_{QCD}$  of the SL QCD
corrections on the non-singlet intercepts are

\begin{equation}\label{slimp}
\epsilon^{(+)}_{QCD} = \frac{\Delta^{(+)}_{QCD} -
\widetilde{\Delta}^{(+)}_{QCD}}{\widetilde{\Delta}^{(+)}_{QCD}}
\approx 11 \%~,\qquad \epsilon^{(-)}_{QCD} =
\frac{\Delta^{(-)}_{QCD} -
\widetilde{\Delta}^{(-)}_{QCD}}{\widetilde{\Delta}^{(-)}_{QCD}}
\approx 9 \%~.
\end{equation}
On the other hand, impacts $\epsilon^{(\pm)}$ of the DL EW
corrections on the DL QCD intercepts
$\widetilde{\Delta}^{(\pm)}_{QCD}$ are of the same size:
\begin{equation}\label{ewimp}
\epsilon^{(+)} = \frac{\Delta^{(+)} -
\widetilde{\Delta}^{(+)}_{QCD}}{\widetilde{\Delta}^{(+)}_{QCD}}
\approx 8 \%~,\qquad \epsilon^{(-)} = \frac{\Delta^{(-)} -
\widetilde{\Delta}^{(-)}_{QCD}}{\widetilde{\Delta}^{(-)}_{QCD}}
\approx -9 \%~.
\end{equation}
Confronting Eq.~(\ref{emimpact}) to Eq.~(\ref{ewimp}) manifests
that the impact of all EW DL corrections on the non-singlet
intercepts is much greater than the impact of the electromagnetic
DL corrections.  It also interesting that EW DL corrections work
opposite ways: they increase $\Delta^{(+)}_{QCD}$ and decrease
$\Delta^{(-)}_{QCD}$, which makes smaller the difference between
the asymptotics of the non-singlets $F_1$ and $g_1$. A qualitative
explanation to that can be easily found from considering
 Eq.~(\ref{intfix}): the expression for $a$ in Eq.~(\ref{adtot})
manifests that adding the EW terms (all they are positive) to the
QCD term $8\alpha_s/3\pi$ increases $a$ and therefore increases
$\omega^{(+)}_0$ compared to its QCD value
$\sqrt{8\alpha_s/3\pi}$. In contrast, there is an interplay
between the increase of $a$ and decrease of $d^{(-)}$ in the
expression for $\omega^{(-)}_0$. Indeed, the QCD term
$8\alpha^2_s/9$ in the expression for $d^{(-)}$ is suppressed by
the negative EW contributions (the largest of them, the second
term, is $\approx -40 \alpha \alpha_s$). It means that
 $\sqrt{a^2 + 4 d^{(-)}} < a$ and
therefore $\omega^{(-)}_0 < \omega^{(+)}_0$.

\section{Conclusion}
We have considered the interplay between the QCD and EW radiative
corrections in the double-logarithmic approximation. We accounted
for the running QCD coupling effects but neglect them for the
electroweak couplings. In the first place we estimated the impact
of EW double logarithms on the amplitudes of the exclusive QCD
processes, considering the annihilation of a quark-antiquark pair
into another quark-antiquark pair of different flavor where the DL
 are known to be  the leading contributions. We found that the EW
impact for the amplitude of this process, though grows fast with
the total energies $\sqrt{s}$, is less than $10 \%$ for $\sqrt{s}
\lesssim 10^3~$GeV. As could be anticipated, the EW impact is more
essential when the total resummation of the DL contributions has
been done than for accounting for the double logarithms in fixed
orders in the couplings.

On the other hand, accounting for the EW DL corrections can lead
to qualitatively new phenomena  which are absent in the QCD
context. As an example, we have considered the EW impact on the
non-singlet structure functions $f^{(\pm)}$ at small $x$ where
accounting for DL contributions is known to be absolutely
necessary. In order to calculate $f^{(\pm)}$ taking into account
both QCD and EW corrections in the DLA, we applied the same method
of composing Infrared Evolution Equations that we had used for
calculating $f^{(\pm)}$ in QCD. The EW couplings to quarks are
sensitive to the quark flavors, so the $Q^2$ and $x$ -evolutions
of $u$ and $d$ -quarks are different. Besides, exchanges with
virtual $W$-bosons mix $u$ and $d$ -quarks. So, accounting for the
EW corrections changes the QCD evolution equation of
Eq.~(\ref{fnsqcd}) for the system of
 more involved equations in Eq.~(\ref{eqfew}).
 Instead of two non-singlet anomalous
dimensions $H^{(\pm)}_{QCD}$ in Eq.~(\ref{hqcd}),
Eq.~(\ref{eqfew}) involves eight of them: $H^{\pm}_{ik}$~, with
$i,k = u,d$. They obey the system of non-linear differential
evolution equations obtained in Eq.~(\ref{eqhik}). The
approximative solutions to Eq.~(\ref{eqhik}) were obtained in
Eqs.~(\ref{hik})~.  They were used to obtain the explicit
expressions of Eq.~(\ref{fpmsa}) for the non-singlet structure
functions $f_u$ and $f_d$ in the kinematic region
Eq.~(\ref{relsqmmu}). Besides, the expressions for $H^{\pm}_{ik}$
in Eq.~(\ref{fpmsa}) can also be used to obtain amplitudes
$M^{\pm}_{ik}$ of the forward annihilation of quark-antiquark
pairs with flavor $i$  into the quark-antiquark pairs with  flavor
$k$: $M^{\pm}_{ik}= 8\pi^2 H^{\pm}_{ik}$~.

In the QCD context, the only difference between the non-singlet
structure functions $f_u$ and $f_d$ is reduced to the difference
in their initial densities $e^2_u \delta u$ and $e^2_d\, \delta
d$~, whereas their coefficient functions and anomalous dimensions
are identical. In contrast, Eqs.~(\ref{fpmsa},\ref{fsa}) manifest
that with the EW corrections taken into account, $f_u - f_d \neq
0$ even if $e^2_u \delta u = e^2_d\, \delta d$~.
Eqs.~(\ref{fpmsa},\ref{fsa}) can also be used for estimating the
$x$ and $Q^2$ -dependence of the asymmetry
\begin{equation}\label{asud}
A_{ud}(x,Q^2)  = \frac{f_u(x,Q^2) -  f_d(x,Q^2)}{f_u(x,Q^2)  +
f_d(x,Q^2)}
\end{equation}
 in the kinematic region
Eq.~(\ref{relsqmmu}). However, the discrepancy between $f_u$ and
$f_d$ does not bring much difference to the small-$x$ asymptotics
of $f_u$ and $f_d$~: they both are of the Regge type with
identical intercepts. Nevertheless, Eq.~(\ref{deltapm})
demonstrates that the EW corrections change the values of the QCD
non-singlet intercepts obtained in Ref.~\cite{egtns} and
reproduced in Eq.~(\ref{deltaqcd}). It is also interesting to
notice that DL contributions of non-ladder Feyman graphs produce
opposite influence on the values of the non-singlet intercepts: In
the QCD framework, the intercept $\Delta^{(+)}_{QCD}$ of the
non-singlet contribution to the structure functions $F_{1,2}$ is
less than the intercept $\Delta^{(-)}_{QCD}$ of the non-singlet
contribution to $g_1$. Eq.~(\ref{deltapm}) shows that accounting
for the EW corrections reverses this situation. Then,
Eqs.~(\ref{deltapm}-\ref{deltaqcddl}) manifest that the impact of
DL EW corrections on the non-singlet intercepts is comparable with
the impact of the sub-leading, i.e. single-logarithmic QCD
contributions and reaches $\approx 11 \%$. As the intercept is the
exponent in the expressions $\sim s^{\Delta}$ for the Regge
asymptotics, the $11 \%$ change of the intercept due to the EW
contributions is quite substantial. Finally, we would like to
stress that similar incorporating EW corrections into the QCD
expressions for the flavor singlet structure functions at small
$x$ should bring really small impact because the small-$x$
behavior of the singlets is mostly controlled by gluon
contributions.

%\section{Figure captions}

\section{Acknowledgements}
We are grateful to R.K.~Ellis who drew our attention to the
problem of interplay between strong and electroweak interactions.
We also grateful to D.A.~Ross for useful remarks concerning the EW
impact on the exclusive QCD processes. The work is partly
supported by the Russian State Grant for Scientific School
RSGSS-5788.2006.2

\end{document}